\documentclass[structabstract]{aa}  
\usepackage{graphicx}
\usepackage{txfonts}
\begin{document}
\def\msun{$M_{\odot}$}
   \title{Stellar Mass and Velocity Functions of Galaxies:\\ Backward evolution \& the fate of Milky Way siblings}


   \author{S. Boissier
          \and
          V. Buat 
	\and
	O. Ilbert
          }


  \institute{Laboratoire d'Astrophysique de Marseille,
     Universit\'e de Provence, CNRS (UMR6110),
38 rue Fr\'ed\'eric Joliot Curie, 13388 Marseille cedex 13, France;
              \email{samuel.boissier@oamp.fr, veronique.buat@oamp.fr, olivier.ilbert@oamp.fr}
             }

   \date{}

 
  \abstract
   {In the recent years, Stellar Mass Functions of both star forming 
and quiescent galaxies have been observed at different redshifts in 
various fields. In addition, Star Formation Rate distributions 
(e.g. in the form of far infrared luminosity functions) were also obtained.
Taken together, they offer complementary pieces of information 
concerning the evolution of galaxies.}
{We attempt in this paper to check the consistency of the observed
  stellar mass functions, SFR functions and the cosmic star formation
  rate density with simple backward evolutionary models.}
   {Starting from observed Stellar Mass Functions for star-forming 
galaxies, we use backwards models 
to predict the evolution of a number of quantities, 
such as the SFR function, the cosmic SFR  density and the Velocity Function. The velocity being a parameter attached to a 
galaxy during its history (contrary to the stellar mass), 
this approach allows us to quantify the number density evolution of galaxies 
of a given velocity, e.g. of the Milky Way siblings. }
   {Observations suggest that the Stellar Mass Function of star forming galaxies is
constant between redshift 0 and 1. In order to reproduce this result,
we must quench star formation in a number of star forming galaxies. The
Stellar Mass Function of these ``quenched'' galaxies
is consistent with available data concerning the increase in the population
of quiescent galaxies in the same redshift interval.
The Stellar Mass Function of quiescent galaxies is then mainly determined by the 
distribution of active galaxies that must stop star formation, with a 
modest mass redistribution during mergers. The cosmic
SFR density, and the evolution of the SFR functions are relatively well 
recovered, although they provide some clue for a small evolution
of the Stellar Mass Function of star forming galaxies at the
lowest redshifts.
We thus consider that we have obtained in a simple way a relatively
consistent picture of the evolution of galaxies at intermediate
redshifts.  We note that if this picture is correct, 50\% of the
Milky-Way sisters (galaxies with the same velocity as our Galaxy, i.e.
220 km/s) have quenched their star formation since redshift 1 (and an
even larger fraction for larger velocities). We discuss the processes
that might be responsible for this transformation.}  
{}


   \maketitle
%

\section{Introduction}

In the last few years, the analysis of many deep fields together with 
improvements in techniques such as  SED-fitting have enabled many
studies of the Stellar Mass Function (SMF) of galaxies at both low and high
redshifts (e.g. Bell et al. 2003, 2007, Borch et al. 2006, 
Bundy et al. 2005, 2006, Arnouts et al. 2007, Vergani et al. 2008, Perez-Gonzalez et al. 2008, 
Pozzetti et al. 2009, Ilbert et al. 2009).
One of the somewhat surprising result is that it is generally found
that the SMF of star forming galaxies (selected on the basis of color,
morphology or spectroscopy) has not evolved much between redshift
about unity and the present time (e.g. Bell et al. 2007, Cowie et al.
2008, Vergani et al. 2008, Pozzetti et al. 2009). The analysis of the
evolution of the Luminosity Function had already suggested that the
stellar mass of blue galaxies is constant, while the one of red
galaxies increases with time (Faber et al. 2007).

Since star formation is undergoing in such galaxies, the stellar mass
in each of them is bound to increase.  This maintained status quo of
the SMF thus indicates that a number of massive star forming galaxies
must evolve towards early type, or at least quiescent galaxies
(through quenching of their star formation), as noted in \cite{bell07}
and \cite{arnouts07}, while they are replaced by lower mass galaxies
gaining mass with time (see also Pozzetti et al. 2009).
This type of scenario was also advanced in Brown et al. (2007)
  to explain the doubling of the stellar mass in red galaxies over the
  past 8 Gyrs.
Such a quenching is also constrained by the evolution of the
co-moving Star Formation Rate (SFR) density, i.e. the SFR
averaged over representative volumes of the universe, dropping by a
factor 6 to 10 between redshift 1 and 0 (see the compilations of
Wilkins et al.  2008 and Hopkins and Beacom 2006).

It is reasonable to think that even at ``intermediate'' redshifts 
(defined broadly as $0<z<1$), 
star formation takes place in ``normal'' star forming galaxies, 
such as the precursors to present day disk-galaxies and the Milky Way.
Smooth SFR histories
are known to be realistic and capture enough of the physics of such galaxies 
(see e.g. models in Boissier \& Prantzos 1999, 2000). 
This is especially suggested by the facts that we see star formation 
at intermediate redshift in rotating disks that should have formed through 
smooth accretion (e.g. Shapiro et al. 2008) and that
most of the galaxies forming the bulk of the infra-red emission at redshift unity look
like undisturbed disk galaxies (e.g. Bell 2005, Melbourne et al. 2005, Lotz et al. 2008).
The tightness of the SFR-Stellar Mass relationship also suggests that the evolution
is mostly driven by smooth evolution and not starbursts (Noeske 2009, Noeske et al.2007).
The kinematics and morphology are more disturbed at high redshift than 
in the local universe but this fact is still consistent with minor merger and gas accretion
(e.g. Ziegler et al. 2009) although this is still debated (e.g. Neichel et al. 2008).
Finally, it has also been shown that models of the chemical and spectro-photometric evolution of the Milky Way and
nearby disk galaxies, when rolled backwards, fit in a very reasonable way the properties
of redshift $\sim$ 0.7 star forming galaxies (Buat et al. 2008).

In this paper, we propose to combine such models with the information from the SMF in order
to compute the evolution of the ``Velocity Function'', with the advantage that the velocity tags a galaxy,
while the stellar mass evolves with redshift when star formation takes place 
(e.g. Noeske 2009  mentions the intrinsic difficulty in comparing 
galaxies of a given  stellar mass at different redshifts). 
This allows
us to check the global consistency of the SMF, the SFR of individual galaxies, 
the cosmic SFR density 
and simple models of galactic evolution. It is then possible to derive 
the fraction of galaxies that were similar to the Milky Way 
(the ``Milky Way siblings'') and that must have quenched their 
star formation during the second half of the history of the universe. 
We acknowledge that this approach is simplistic but we think it is worth 
performing it since it allows us to step back and have a global look at
all the important data of the problem (Stellar Mass Functions of quiescent
and star forming galaxies, SFR functions, cosmic SFR  density) in a simple, educational 
framework, while many studies focus only on some of these aspects (with however 
interesting exceptions as e.g. Hopkins et al. 2009 or Bell et al. 2007).

In section \ref{secModels}, we present the ``backward'' models of the evolution
of star forming galaxies that we use. In section \ref{secVel}, we
compute the velocity function and its evolution assuming a constant 
SMF, as suggested by the observations. 
In section \ref{secEvo}, the consequences of our simple assumptions are discussed. In section
\ref{SecConclu}, a summary of our work is proposed.


\section{Backward evolutionary models}
\label{secModels}

\begin{figure}
  \centering
  \includegraphics[width=9cm]{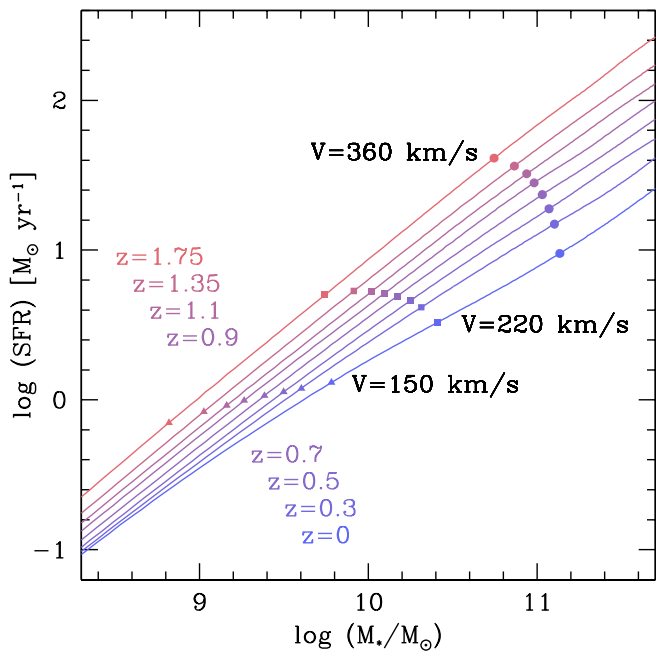}
  \caption{Redshift evolution of the SFR - Stellar Mass relationship in the backward 
          models. Symbols (triangles, squares, circles) show the values for three velocities
 	(150, 220, 360 km/s).}
  \label{FigModelsSFRvsM}
\end{figure}

\begin{figure}
  \centering
  \includegraphics[width=9cm]{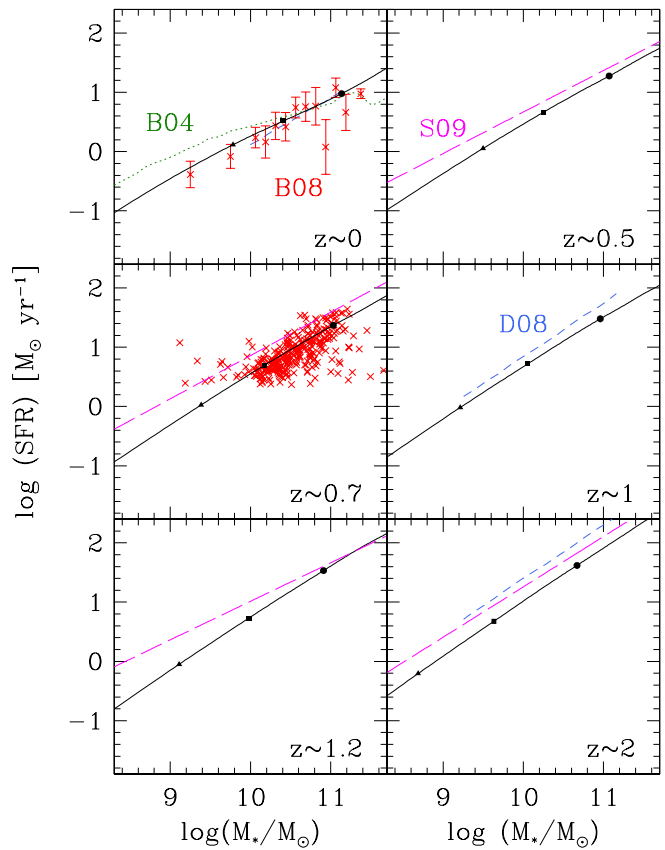}
  \caption{Observed SFR - Stellar Mass relationship at various
    redshifts, compared to the models (solid line with the same symbols as in 
Fig. \ref{FigModelsSFRvsM}). Blue dashed lines (D08): 
observational trends reported by \cite{dave08} on the basis of \cite{elbaz07} and
\cite{daddi07}. Dotted line (B04): \cite{brinchmann04}. Crosses (B08): \cite{buat08}. 
    Magenta Long-Dashed lines (S09): \cite{santini09}.}
  \label{FigDataSFRvsM}
\end{figure}

Following the ``backward'' approach of \cite{silk99},
we make the assumption that the properties of star
forming galaxies at high redshift can be inferred by rolling backward
models of the evolution of well studied present day disk galaxies.
This is motivated especially by \cite{buat08}, showing 
that such backward models reproduce the average specific SFR - 
Stellar Mass relationships observed in galaxies at redshift 0 and 0.7.

In the present work, we use the same basic models and we remind below
their main ingredients.  These models are based on the grid proposed
in \cite{boissier99} and \cite{boissier00} for the Milky Way and
nearby spiral galaxies, calibrated to reproduce many observed trends
in the local universe (e.g. colour-magnitudes diagrams,
luminosity-metallicity relationship, gas fractions, 
colour and metallicity gradients). A
comparison of this grid with the properties of galaxies at redshifts
lower than 1 was proposed in \cite{boissier01}.
Note that we changed the Initial Mass Function (IMF) from
\cite{kroupa93} to \cite{kroupa01}, motivated by the fact that this
IMF is closer to the one used in general in studies of the SMF. It is
also providing a better fit to the UV surface brightness profiles of some of
the SINGS galaxies (Mu\~nos Mateos et al., in preparation). The change
affects mostly the most massive stars (thus the UV emission and
metallicity output), leaving the SFR, gas, stellar mass evolution
changed by less than about 20\%. This number is below the
  typical uncertainty that we can expect in our models. Indeed the
  calibrations used to measure stellar masses or SFR on which are
  based the models can easily be changed by a factor 1.5 to 2, for
  instance adopting different IMFs (e.g. Bell et al. 2007). The
  efficiency of the star formation law, a basic ingredient of the
  model is also varying within a factor 2 between galaxies (e.g.
  Kennicutt et al. 1998). As a result, we do not expect such models
  to give results to a better accuracy than a factor 2.

The grid of models was constructed adopting various circular 
velocity ($V$)  and various values of the spin parameter 
($\lambda$) measuring the specific angular momentum of the
galaxy. With simple scaling relationships, this allowed
the reproduction of the observed ranges of  stellar masses and sizes of disk galaxies 
(the used scaling relationships imply that galaxies with larger 
velocities are more massive, and galaxies with larger spin 
parameter have larger spatial extent).
For each couple of values of the parameters ($\lambda,V$), the 
corresponding model provides one unique history of the galaxy that
includes the evolution of the SFR and of the stellar mass, resulting 
from the combination of a star formation law (see e.g. 
Kennicutt 1998, Boissier et al. 2003) and the accretion of cold gas 
at a rate choosen to reproduce the observations in nearby galaxies.
For simplicity and following \cite{buat08}, we adopt here the average
value of the spin parameter ($\lambda$=0.05), and concentrate on the
dependences on  stellar mass (the mass of a galaxy depends on the
circular velocity as $M \propto V^3$). A distribution of spin
parameter would create a small dispersion around the main trends shown
in the paper as a function of the stellar mass.
The fact that the mass is the main intrinsic parameter of the evolution of galaxies was already 
found in ``local'' studies (e.g. Gavazzi et al. 1996) and confirmed in the recent years 
at high redshift (e.g. Noeske 2009).
Thus, in this paper, for a given velocity, we have one star formation history,
that can be seen in Fig. 6 of \cite{buat08}. 
For high velocities, the SFR 
presents a rapid rise followed by a exponentially 
declining star formation rate. For $V=220$ km/s
(corresponding to the case of our Milky Way) 
the history is similar with a modest
decline (in agreement with the observational constraints available in 
our Galaxy). 
Going to even smaller velocity (towards less massive galaxies), the SFR
gently rises with time. Summarizing in simple words, the stars 
are formed earlier in galaxies with large velocities than in those with 
low velocities.
We stress that these SFR Histories are not especially picked for this work. 
On the contrary, they result from the assumptions that one has to make to reproduce
among others the distribution of metallicity in the Solar Neighborhood G-dwarf stars
(Boissier \& Prantzos, 1999) implying a long time-scale for the formation
of our Galaxy, the colors and metallicity (Boissier \& Prantzos 2000), 
the star formation efficiency and gas fractions (Boissier et al. 2001) 
of nearby galaxies, especially as a function of the luminosity or
the stellar mass. Of course, the evolution might be more complicated 
than in  these models, and there is no unique solution, however the 
general trends 
of these star formation histories are robusts in view of the
constraints involved.

Since Stellar Mass and SFR play a central role in our analysis, we show
in Fig. \ref{FigModelsSFRvsM} the relation between these two quantities
in our models for various redshifts. The
redshifts are chosen to correspond to the data of \cite{ilbert09} and
\cite{arnouts07} to which we compare our results later in the paper.
Note that the dependence of the various quantities on the circular velocity were
initially computed only for a few points. These relations were interpolated on a finer grid
and smoothed in order to avoid oscillations or artificial breaks due to the limited number
of computed points. This allows us to present a smooth curve 
in this plot and the others in the remaining of the paper.

As mentioned earlier, there are evidences that these simple models
capture enough of the properties of star forming galaxies up to
redshift $\sim$ 0.7 (Boissier \& Prantzos 2001, Buat et al. 2008). 
We present in Fig. \ref{FigDataSFRvsM} in addition the relationship
between the stellar mass and SFR for a set of redshifts for which
various observations are available. This simple comparison shows that
up to redshift 1, the model reproduces quite well the average trends
found in observations, with serious departures starting only to be
present at redshifts larger than 1.  Especially at redshift $\sim$ 2,
for galaxies more massive than 10$^{10}$ \msun (for which observations
are available), the models tend to underestimate the actual SFR for a
given stellar mass. This difference between observations and
models was found by other authors (e.g. Dav\'e 2008, Santini et al.
2009) using semi-analytical or hydrodynamical models.  
\cite{dave08} suggests that a
solution might be a modification of the IMF with redshift.  As noted
by \cite{santini09}, \cite{khochfar09} propose that the star formation
efficiency in high redshift galaxies is higher owing to cold gas
accretion. Neither of these effects (probably still to be debated) are
included in our models, and our strongest conclusions will be limited
to redshifts lower than about unity, where they are not needed.

The various quantities presented in the paper assume a \cite{kroupa01} IMF, and a 
($H_0$=70, $\Omega_{M}$=0.3, $\Omega_{\lambda}$=0.7) cosmology.

\section{Derivation of the velocity function}

\label{secVel}

\begin{figure}
  \centering
  \includegraphics[width=9cm]{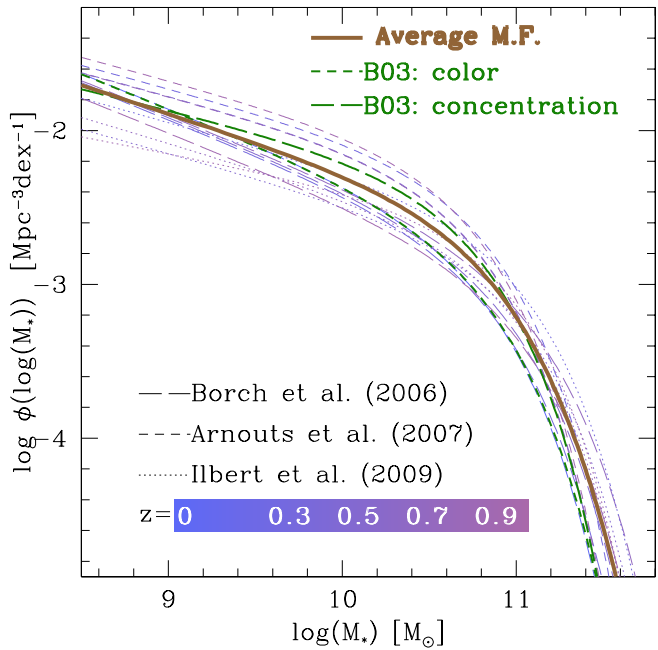}
  \caption{Stellar Mass Functions for star forming galaxies
from \cite{bell03} (redshift 0, bold dashed curves for
both the color-selected sample and the concentration-selected one), \cite{ilbert09} as dotted curves, \cite{arnouts07} as short-dashed curves and
\cite{borch06} as long-dashed curves (between redshifts 0 and 1). The resulting Average Mass 
Function (AMF) is shown as the bold solid curve.}
  \label{FigAverageMF}
\end{figure}

The SMF of star forming galaxies has little evolved between redshift 1
and redshift 0. In Fig. \ref{FigAverageMF}, we present the
observed SMF of \cite{bell03} at redshift 0 (with two
different definitions of the star forming sample, based on either
color or concentration); and those from \cite{ilbert09},
\cite{arnouts07}, and \cite{borch06} in various redshift bins
(restricted to z$<$1). This figure confirms that the evolution of the
SMF for Star Forming galaxies, if any, is quite small (lower
  than 0.2 dex) with respect to the difference between the
various studies that can reach about 0.5 dex.

Taking into account this dispersion and lack of clear evolution, we
believe it is educational to adopt the assumption that the SMF is
indeed constant with redshift, by defining an Average Mass Function
(AMF) as the average value among the curves shown in Fig.
\ref{FigAverageMF} (unweighted average of log($\Phi(M_*)$
  after converting each work to the Kroupa et al. 2001 IMF).
This AMF is shown in the figure as the thick curve.  Note that many
other determinations of the SMF exist, and we limited ourselves to a
few representative cases.  Especially, \cite{bell03} is often used as
a local reference, and the SMF by \cite{borch06} were measured following
the same methods at higher redshifts. \cite{ilbert09} and
\cite{arnouts07} propose two slightly different approaches in
identical redshift bins, and procure a good example of the dispersion
that can be found in the data due to e.g. cosmic variance or
differences in the adopted methods. 
Note that \cite{hopkins09} also choosed these two last studies for
their analysis for the same reasons.  Inspection of Fig.
\ref{FigAverageMF} thus shows that the AMF at each stellar mass has a
typical uncertainty of 0.2 dex.

\begin{figure}
  \centering
  \includegraphics[width=9cm]{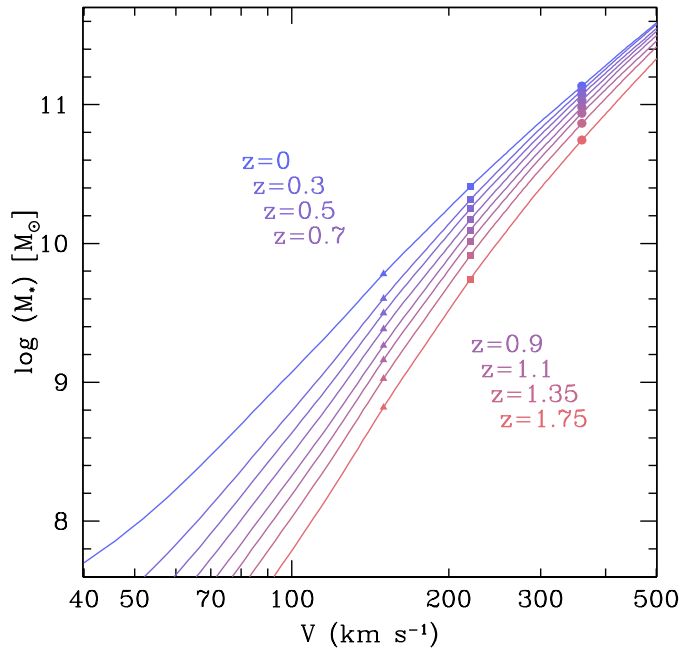}
  \caption{Stellar mass - velocity relationship in the models 
(same color and symbol-coding as in Fig.  \ref{FigModelsSFRvsM}).}
  \label{FigMassV}
\end{figure}

As stellar mass is obviously evolving in star forming galaxies, the
fact that the SMF is roughly constant for $0<z<1$ is actually hiding a
real evolution.  To get rid from this difficulty, we compute a model
Velocity Function where the velocity is actually the parameter of the
simple model presented in section \ref{secModels}, and thus does not
evolve. The Velocity Function is computed as:
\begin{equation}
\phi(V)=\frac{dN}{dV} = AMF(M) \times \frac{dM}{dV}.
\end{equation}
In this equation, AMF(M) is the Average Mass Function defined above,
and the factor $dM/dV$ is obtained for each redshift by numerically
deriving the velocity- stellar mass relationships of the models shown in Fig.
\ref{FigMassV}: for each redshift, there is a one-to-one
correspondence between the stellar mass and the velocity.

\begin{figure}
  \centering
  \includegraphics[width=9cm]{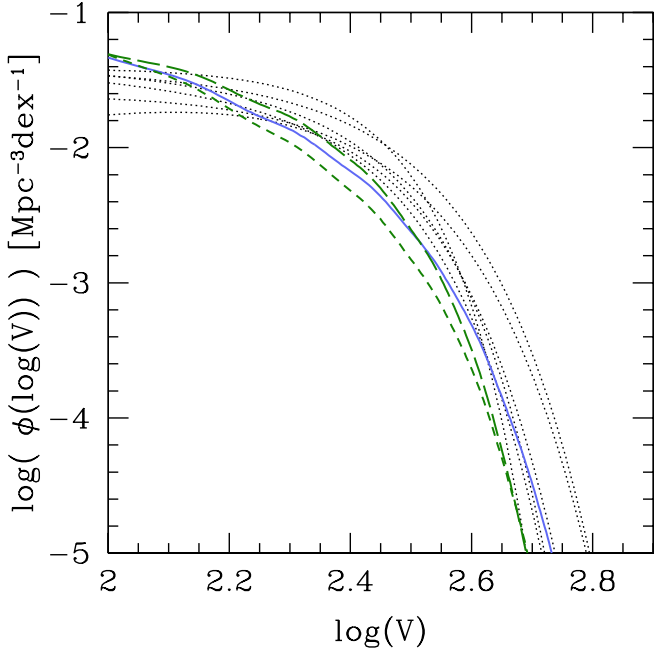}
  \caption{Spiral galaxies Velocity Function at redshift zero,
      derived from the AMF using our models (solid line), compared to
      velocity functions empirically derived in the local universe by
      \cite{gonzalez00}, using various observational sets (dotted
      lines). The dashed curves show the Velocity Functions obtained
      if the Bell et al. (2003) redshift 0 functions (shown in Fig.
      \ref{FigAverageMF}) are used instead of the AMF.}
  \label{FigVfuncZ0}
\end{figure}

By definition, at redshift 0, the circular velocity can be identified
with the observed rotational velocity. We thus compare in Fig.
\ref{FigVfuncZ0} the Velocity Function obtained for redshift 0 with
the rotational velocity functions for spirals observationally
derived by \cite{gonzalez00}. They obtained their functions by
combining various empirical luminosity functions and tully-fisher
relationships available at that time.  Taking into account the scatter
they obtained for different inputs (about 0.4 dex at V=220
  km/s), and the scatter around the AMF (different studies
  and different redshifts cover a range of 0.5 dex ), the agreement
between our redshift 0 model Velocity Function and the ones
in \cite{gonzalez00} is reasonable.  We also show the Velocity
Functions obtained if we replace the AMF by the redshift 0
observed Stellar Mass Functions from \cite{bell03}, giving an
idea of the uncertainty (the scatter in Fig.  \ref{FigAverageMF} is
larger than the difference obtained using the AMF or Bell et al.
functions)

\section{Cosmic evolutions}
\label{secEvo}

\subsection{Evolution of the velocity and SFR functions}
\label{secSfrFunc}

\begin{figure}
  \centering
  \includegraphics[width=9cm]{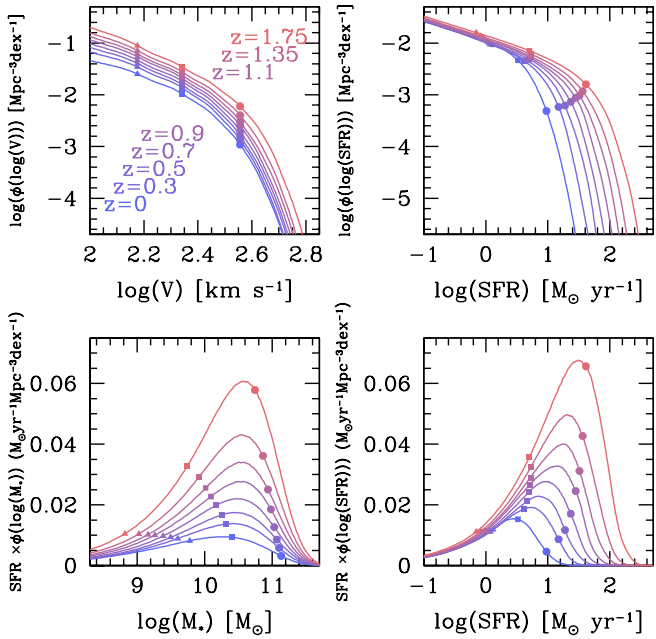}
  \caption{Redshift evolution of the modeled velocity distribution (top-left
    panel), SFR function (top-right panel) and contribution
    to the cosmic SFR density (bottom) of galaxies of a given
    stellar mass (left) or SFR (right), assuming a constant
    stellar Mass function (the AMF, shown in Fig.
    \ref{FigAverageMF}). The redshifts and symbols are the same as in
    Fig. \ref{FigModelsSFRvsM}.  }
  \label{FigMultiEvo}
\end{figure}

In the previous section, we derived from the combination of
observations (a constant stellar mass function, the AMF), and  the stellar mass-velocity 
relationship of simple models the velocity function at various
redshifts. 
From this distribution at each redshift, many quantities
and their distribution can be computed using the models. 
Fig. \ref{FigMultiEvo} shows the evolution of the model 
velocity function in the top
left panel. The ``SFR function'' evolution is shown in the
top right panel, and the bottom panels shows the evolution of the
contribution to the cosmic SFR  density  by galaxies of various stellar mass
(left) and SFR (right). 
These panels illustrate that at higher redshift, the cosmic SFR
density is dominated by galaxies of larger 
SFR in our models.
It is also dominated by galaxies of larger stellar masses at
larger redshift, although the effect is smaller.
The peak of the contribution to the cosmic SFR
density moves from $log(M/M_{\odot}) \sim 10.6$ (redshift
1.75) to $log(M/M_{\odot}) \sim 10.3$ (redshift 0).  In the following,
we study some interesting details of the evolution summarized in this
figure.

\begin{figure}
  \centering
  \includegraphics[width=9cm]{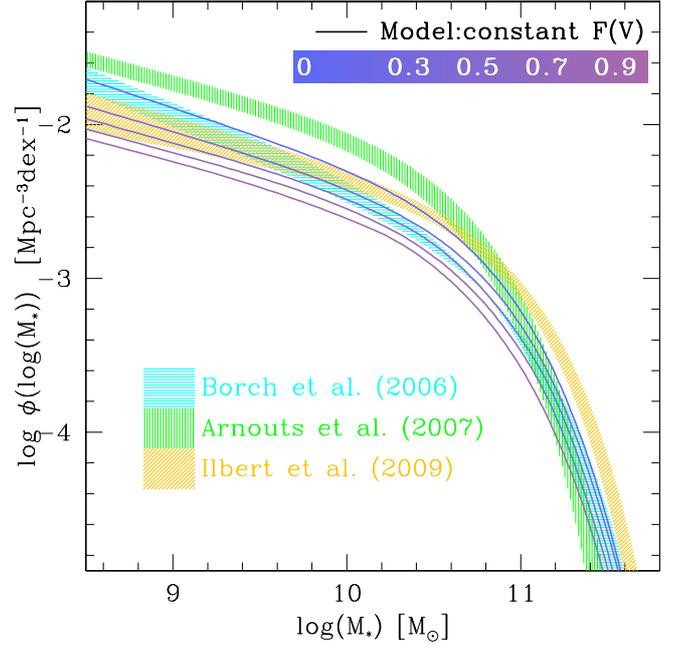}
  \caption{Predictions at redshifts 0, 0.3, 0.5, 0.7, 0.9 for the
    Stellar Mass Function of star forming galaxies assuming a constant
    Velocity Function (lines), compared to the observed ranges by
    \cite{ilbert09}, \cite{arnouts07} and \cite{borch06} in the same
    redshift intervals.}
  \label{FigCstFV}
\end{figure}

These results (and those in the remaining of the paper) have
  been obtained assuming that the Stellar Mass Function of star
  forming galaxies is constant with time, while their Velocity
  Function changes.  Before studying in details the consequences of
  our assumption, we first try to see wether the opposite
  assumption is acceptable. We thus assume that the Velocity Function
  is constant (we adopt the velocity function found at redshift 0
  above, and shown in Fig. \ref{FigVfuncZ0}), and compute the
  evolution of the Stellar Mass Function then implied by the model
  stellar mass-velocity relationship.  The results for redshift lower than
  1 are shown in Fig.  \ref{FigCstFV}, together with the area occupied
  by the Star Forming Galaxies Stellar Mass functions observed in the
  same redshift interval by \cite{ilbert09}, \cite{arnouts07} and
  \cite{borch06}. We find a systematic decrease of the computed
  Stellar Mass Function with redshift.  For
  $log(M/M_{\odot})=10.5-11$, the decrease with respect to the
  redshift 0 value reaches 0.4 dex at redshift $\sim$ 1.  This number is
  not very far from the dispersion found among the observational studies. 
  However, this dispersion is largely due to systematic
  differences between the various studies.  Whithin the individual
  studies of \cite{ilbert09}, \cite{arnouts07}, and \cite{borch06},
  Fig \ref{FigCstFV} shows that the evolution of the SMF is smaller
  than $\sim$ 0.2 dex, a much lower value than the change we expect
  under the assumption of a constant Velocity Function.  We thus
  conclude that the evolution of the star-forming galaxies Velocity
  Function is clearly established.

\subsection{Build-up of the ``quiescent''  stellar mass function}

\label{secbuiltup}

\begin{figure}
  \centering
  \includegraphics[width=9cm]{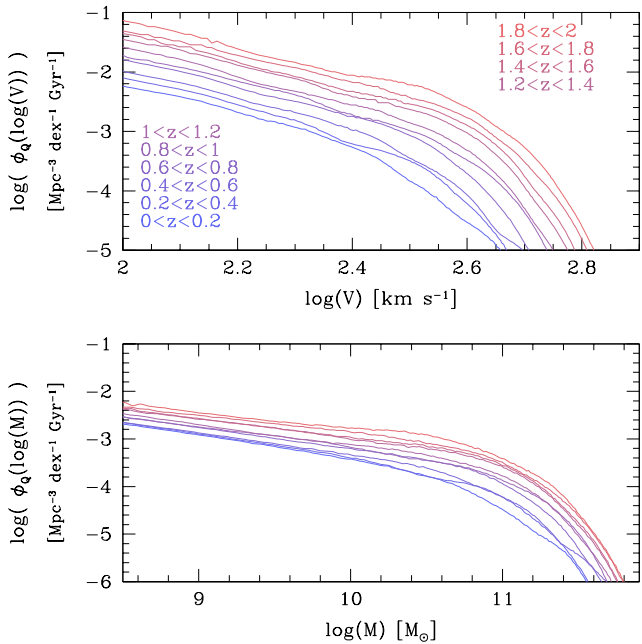}
  \caption{Velocity (top) and stellar Mass Functions (bottom)
    of galaxies that ``Quenched'' their star formation within
    redshift bins of $\Delta z$=   0.2,  as predicted in our model. The
    functions are normalised by the time elapsed within the
    corresponding redshift bins so that they correspond to
    number density fluxes.}
  \label{FigQuenchMF}
\end{figure}

\begin{figure}
  \centering
  \includegraphics[width=9cm]{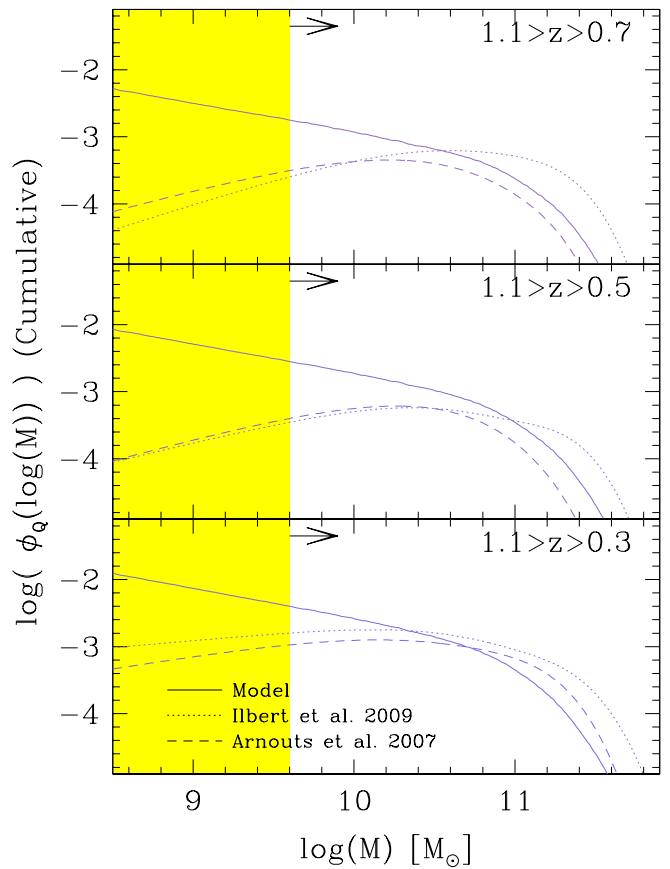}
  \caption{The cumulated quenched stellar mass function since
    z=1.1 in our simple model is compared to the increase of the
    observed stellar mass function of quiescent galaxies in
    \cite{ilbert09} and \cite{arnouts07}. The comparison is meaningful
    only out of the shaded area, on the side indicated by the arrow
    (the limit corresponds to log(M)=9.6, the lower stellar
    mass used in \cite{ilbert09} at redshift 1.1, that we use as
    reference.}  \label{FigQuenchCumul} \end{figure}
\begin{figure*}
  \centering
  \includegraphics[angle=-90,width=17cm]{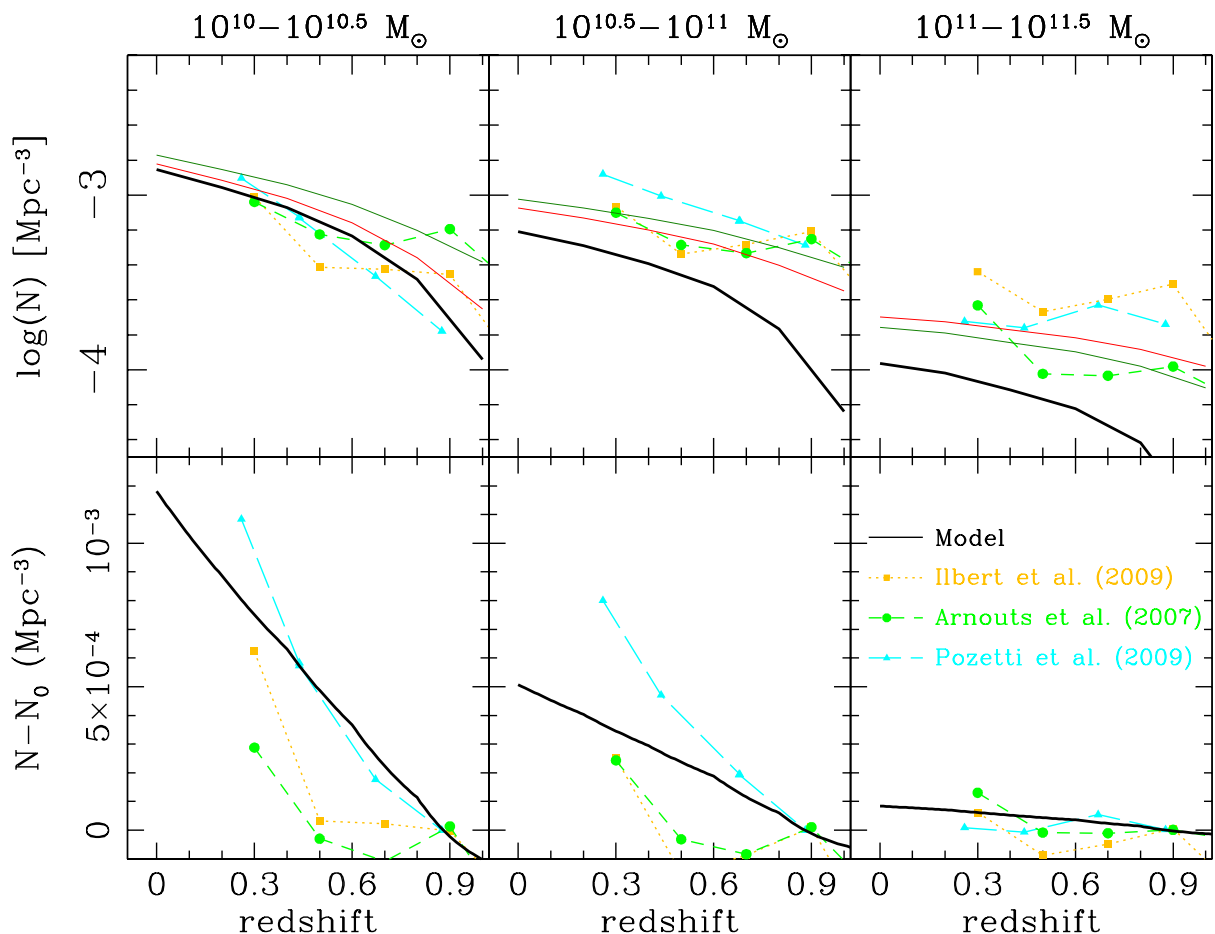}
  \caption{\emph{Top:} Modeled evolution of the number density of
    quiescent galaxies (solid curves) in three stellar mass bins (left
    to right: 10$^{10-10.5}$ \msun, 10$^{10.5-11}$ \msun{} and
    10$^{11-11.5}$ \msun) cumulated since redshift 1.1, compared to
    the data of \cite{pozzetti09} for the same mass ranges as the long
    dashed curves with triangles. We integrated within the same mass
    bins the observed stellar mass functions of
    \cite{ilbert09} and \cite{arnouts07}, respectively the dotted with
    squares and dashed with circles curves). We also show as two thin
    solid curves how the model curve is shifted if we add to the newly
    quenched galaxies the number of quiescent galaxies already present
    at z=1.1 according to the observed Stellar Mass Functions
    of \cite{ilbert09} and \cite{arnouts07}.  \emph{Bottom:}
    Differential number density evolution in the same mass ranges
    ($N_0$ is the number density at redshift 0.9 in both the data and
    the model). See section \ref{secbuiltup} for details.}
  \label{FigCumulDens}
\end{figure*}

 The
model velocity function for star-forming galaxies
shown in Fig. \ref{FigMultiEvo} evolves with redshift, with a larger
number of star forming galaxies at higher redshift for any given
velocity. It is another way to say that, at a given velocity, a
fraction of galaxies have quenched their star formation between the
redshift probed and now.  We computed the velocity function with a
$\Delta z$=0.2 step in redshift. The difference between functions
at different steps is the velocity function of galaxies that have quenched
their star formation rate in this redshift interval and is shown in
the top panel of Fig. \ref{FigQuenchMF} (normalised to be expressed as
a flux).  Combining it with the stellar Mass-Velocity
relation at the beginning of each redshift interval (assuming the
quenching is instantaneous at the beginning of the bin, since the
stellar mass evolves little in such small redshift interval),
we can compute the Quenched stellar Mass Function in each
interval (bottom panel).  This function is normalized by the time
elapsed within each redshift bin, so that it is actually a
number density flux from star forming galaxies to quiescent
ones. These numbers can be compared to e.g. \cite{pozzetti09} who
found a red sequence growth rate of a few 10$^{-4}$ gal Mpc$^{-3}$
Gyr$^{-1}$ dex$^{-1}$ for $log (M/M_{\odot}) <$ 11 in the redshift
range 0.3 to 0.9.  With the quenched flux estimated above, it is easy
to compute the cumulated Quenched stellar Mass Function
between two redshifts.  This is done in Fig. \ref{FigQuenchCumul},
between redshift 1.1 (beyond this point, the models start to be
unreliable, see section 2) and several redshifts for which we compare
the results of our model to the increase in the quiescent
stellar mass function of \cite{ilbert09} and
\cite{arnouts07}.  Considering the uncertainties in the derivation of
stellar mass functions (especially at high redshift, see in
the figure the difference between the two observational studies), the
agreement is good above $\sim$ 3 10$^{10}$ \msun, i.e. the
increase in the observed stellar mass function of quiescent
galaxies is well compatible with the number of star forming galaxies
that must quench their star formation in order to obtain their about
constant stellar mass function, within a few tens of
  dex, i.e.  our uncertainties (see also Bell et al. 2007 for a
simple approach, and Pozzetti et al 2009).  Below and around 10$^{10}$
\msun, the model predicts a much larger number of quenched galaxies
and a steeper slope than the differential evolution obtained in the
observational studies.  This results directly from the assumption that
the SMF of star forming galaxies does not evolve, even at low
stellar mass where it is not observed at high redshift (in
this regime, the model is a mere extrapolation). To obtain less
quiescent galaxies with stellar masses in the range 10$^9$ -
10$^{10}$ \msun, we would need a change in the number of low-mass star
forming galaxies with redshift. Only future observations bringing
constraints on the low-mass end of the stellar mass function
of star forming galaxies at intermediate redshift will allow us to
check if it is the case.  \cite{pozzetti09} computed the number
evolution of quiescent galaxies in three stellar mass ranges
between redshifts $\sim$ 0.25 and 0.9. In Fig. \ref{FigCumulDens}, we
show their data and compare them to the predictions of our model
(solid curves, integrating the cumulative quenched stellar
mass function such as the ones shown in Fig. \ref{FigQuenchCumul} in
the same stellar mass bins). We also show our computations of
the same numbers using the observed stellar Mass Functions of
\cite{arnouts07} and \cite{ilbert09}, allowing to estimate the
uncertainty on these numbers by comparing the various studies.  When
considering the top panels, one has to take into consideration the
fact that a number of quiescent galaxies are ``formed'' beyond
redshift 1.1, while the model only shows the one formed by quenching
after this epoch. For this reason, the two thin lines show the sum of
the newly quenched galaxies (model) and of the quiescent galaxies
already existing at redshift 1.1, taking the data of \cite{ilbert09}
or \cite{arnouts07}.  We also show in the bottom panels the relative
number evolution with respect to $z \sim$ 0.9.  On the massive
galaxies side, the quenching forms a small number of quiescent
galaxies with respect to the number already present at z=0.9: massive
galaxies are already in place at this redshift. The model predict a
small increase of the number density in this stellar mass
range, consistent with the estimations of a quasi constant
number-density.  Moving to the intermediate stellar mass bin
(10$^{10.5-11}$ \msun{}), we predict a larger increase in the number
of quenched galaxies, consistent with the data. We note however a
large dispersion: the model tend to underpredict the evolution with
respect to \cite{pozzetti09}, but overpredict it with respect to
\cite{ilbert09} or \cite{arnouts07}.  In the lower mass bin, the model
is roughly consistent with the various studies when no
  normalisation is done (top panel of fig.  \ref{FigCumulDens}). In
  the bottom panel, two of the observational works are found below
  the model curve but this is mostly due to the z=0.9 point.
The figure suggests that most of the low-mass quiescent
galaxies are formed between redshifts 1.1 and 0 through quenching of
the star formation of star-forming galaxies.

\subsection{On the role of mergers}

\begin{figure}
  \centering
  \includegraphics[width=9cm]{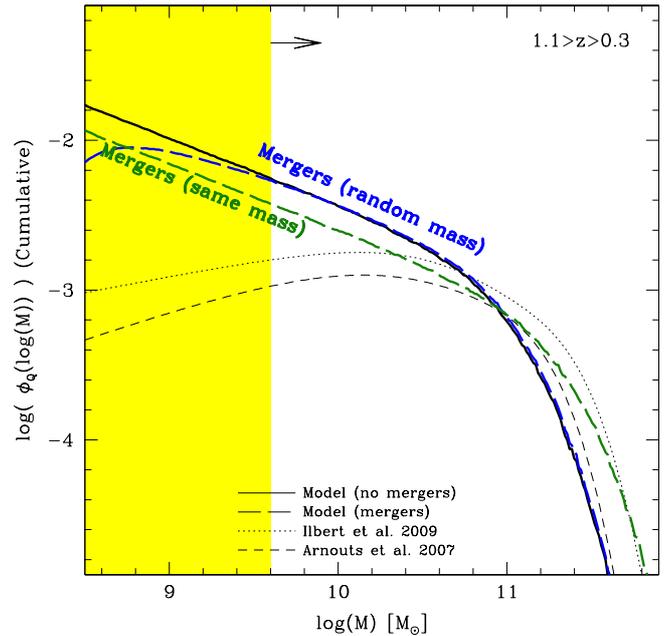}
  \caption{The increase of the observed stellar mass function of
    quiescent galaxies in \cite{ilbert09} and \cite{arnouts07} is
    compared to monte-carlo realization of the cumulated quenched mass
    function between z=1.1 and z=0.3. One realisation (black solid
    line) shows the same as Fig. \ref{FigQuenchCumul} (i.e. result
    obtained by quenching star-forming galaxies by the right amount to
    keep a constant  stellar Mass Function) while others show the potential
    effect of various merger scenario on the mass function (see
    text).}
  \label{FigMerger}
\end{figure}

Fig. \ref{FigQuenchCumul} shows that the star formation quenching
inferred from the model reproduces relatively well the
observed increase on the massive side of the SMF for
quiescent galaxies (even if uncertainties due to poor statistics are
large). It overpredicts the quiescent SMF on the low mass-side (even
if at the lowest stellar masses, the models are
extrapolations, the slope seems larger than the one observationaly
found). How would those results be affected if in addition to the
quenching itself, the mass is re-distributed through mergers ?
A complete answer would need a full treatment of the 
Halo Occupation Distribution and is beyond the scope of this paper.
However, we attempt to illustrate what could happen in two simple cases
shown in Fig. \ref{FigMerger}.
We thus start from the modeled Stellar Mass Function of
quenched galaxies between redshifts 1.1 and 0.3 (within the mass range
$10^8$ to $10^{12}$ $M_{\odot}$).  We then assume two merging
scenarios, redistributing the mass in different ways: - Case 1: Each
quenched galaxy merge with another one, the mass of each merging
galaxy being taken randomly from the initial distribution with
  lower mass limit $10^8 M_{\odot}$ (mergers - random mass curve in
Fig. \ref{FigMerger}), resulting in an average mass ratio of 0.14 for
a final 10$^{11}$ \msun{} galaxy.  - Case 2: Each galaxy suffers one
major merger in the redshift interval, computed by assuming that each
galaxy merge with a galaxy of identical mass (same mass curve).
\cite{jogee09} estimated that 68 \% of galaxies with stellar
masses larger than 2.5 10$^{10}$ \msun{} have undergone a merger of
mass ratio larger than 1/10 over lookback times of 3 to 7 Gyrs.
Including stellar masses as low as 10$^9$ \msun, they found
that 84 \% of galaxies have undergone one merger.  Thus Case 1 may not
be very far from the truth.  Case 2 is highly biased in favor of major
mergers given the results from various studies. For instance,
  \cite{jogee09} found that 16 \% of galaxies have undergone a major
  merger over lookback times 3-7 Gyrs.  Based on galaxy pairs,
  \cite{deravel09} found that 22 \% of galaxies have undergone a major
  merger since z $\sim$ 0.9.  \cite{carlos09} found that 8 \% of
  galaxies more massive than $10^{10} M_{\odot}$ have undergone a
  merger since z $\sim$ 1.

Case 1 shows that a small number of minor mergers have almost
no effect on the  stellar  mass function. Making e.g. $\sim$ ten minor 
mergers (an extreme case) rather than 1 would have an impact 
on the results however they would depend on the lower mass limit adopted 
for the computation, while the SMF is increasingly 
uncertain towards low masses.

Case 2 shows what happens if each galaxy goes through one major
merger. In this case, the shape of the quenched stellar Mass
Function is modestly changed at low and intermediate mass where the
number of galaxy at a given stellar mass is slightly reduced.
Only the high-mass end of the distribution is significantly changed.
In particular, major mergers seems to be the only way to build the more
massive quiescent galaxies.

As seen earlier, the slope found in the intermediate/low mass 
range of the diagram is quite steep in the model with respect to the observations of 
\cite{ilbert09} or \cite{arnouts07}. Fig. \ref{FigMerger} shows that
simple merger models do not help to solve the problem.

One assumption that has been implicitly done is that mergers do not
enhance star formation. If it was not the case, a fraction of
  star forming galaxies woud have larger SFR than deduced from the
  models, and the result of the fusion would have a larger stellar
  mass than the one described in the basic approach of this section.
However, \cite{robaina09}
showed recently that the additional SFR due to mergers is 
very modest (and then even more significant for the additional stellar mass). 
\cite{deravel09} found a net star formation enhancement in merging 
pairs of only 25 \%.
This modest enhancement is also consistent with the merger simulations
of \cite{dimatteo07}.
Note that in fact, we do not exclude at this level the possibility 
of mergers being related to the quenching itself. The merging
of two star forming galaxies of similar mass may result
in the quenching of their star formation, for instance through AGN 
formation and feedback (e.g. Sanders et al. 1988,
Springel et al. 2005). This case would be exactly similar 
in our approach to two star-forming galaxies quenching their star 
formation, and having their masses added together to form a new galaxy (Case 2 above).

In summary, in this section, we only discussed about the shape of the
mass distribution of quiescent galaxies, and the possible impact of
mergers on it.  Our conclusion is that the stellar Mass
Function of galaxies quenched at redshift lower than 1.1 
originates from the shape of the SMF of star forming galaxies.  Mergers only
bring modest adjustment to the quenched stellar Mass Function
by re-distributing part of the mass.  They probably may play an
important role at most at the high-mass end of the Mass Function.

\subsection{Cosmic SFR density}

In the last decade, many studies have put constraints on the cosmic
SFR density history. A compilation can be found in
\cite{hopkins06}.  In a more recent work, \cite{wilkins08} derived the
cosmic SFR density  history from a compilation of measurements of the cosmic
stellar density at various redshifts. These two approaches give close
results in the redshift range in which we are interested, the latter
one producing a smaller drop in the cosmic SFR density  between redshift 1 and
0 (a factor of about 6 versus 10).  We decided to use as a reference
the results of \cite{wilkins08} for two main reasons: first it is
based on measurements of mass densities, more consistent with our
starting points of  stellar mass functions than SFR measurements.  
Second, 
any evolution that we need to impose in our 
model to reproduce this trend is likely to be a lower
limit on the actual evolution.

\begin{figure}
  \centering
  \includegraphics[width=9cm]{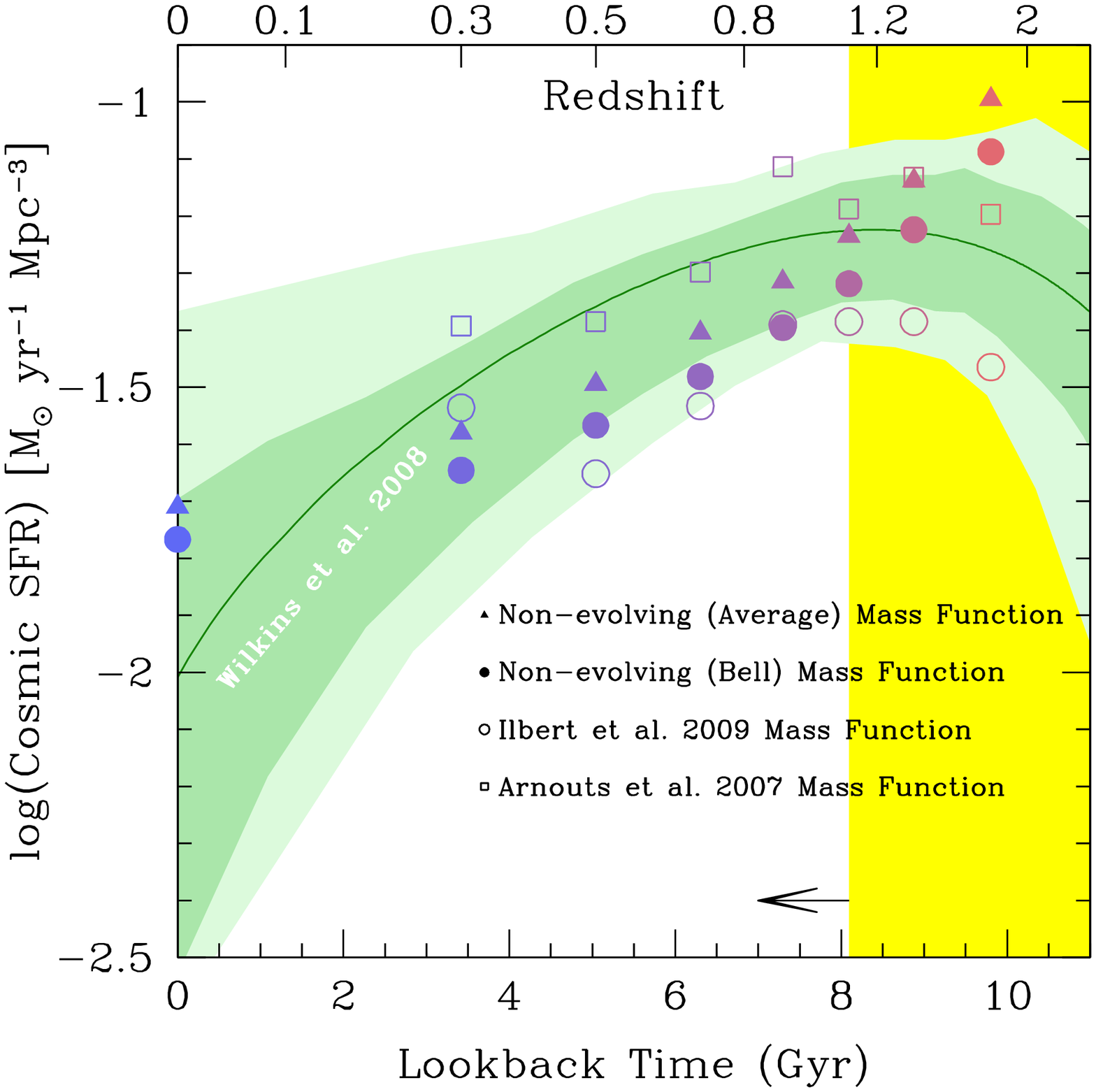}
  \caption{Cosmic star formation rate density histories. The
    observational trend from \cite{wilkins08} is compared to model
    predictions assuming a constant SMF (filled circles assuming the
    \cite{bell03} stellar Mass Function for a color-selected
    sample, filled triangles assuming the AMF), or using for each
    redshift the SMF observed at this redshift by \cite{ilbert09}
    (open circles) or \cite{arnouts07} (open squares).  In the
    vertically shaded area ($z>1.1$), the results from the models are
    mere extrapolations showing the limits of our approach beyond this
    redshift.}
  \label{FigCosmicSFR}
\end{figure}

In Fig. \ref{FigCosmicSFR}, we reproduce the evolution provided by
\cite{wilkins08} with their 1 and 3 $\sigma$ dispersion determination
as the solid curve and shaded areas.  We overplot the cosmic history
of the SFR density obtained with the model velocity
function derived at various redshift as explained before, in
combination with the velocity-SFR relation from the models
(triangles).
Concentrating on the redshifts lower than 1 where we are confident in
the models, we note that the points are within the shaded area
determining the compilation of data from \cite{wilkins08} even if the
drop in SFR is not as large as usually considered (factor of $\sim$ 3
rather than 6 since redshift 1).  This is mostly due to the fact that
our AMF produces a relatively large cosmic SFR density at
redshift 0. We also present in the figure the cosmic SFR
density if the observed Stellar Mass Function of
\cite{bell03}, derived at redshift zero is used rather than the AMF
(dots). Very similar results are found, especially the fact that our
redshift 0 point is relatively high. This indicates a small
  inconsistency between the SMF of local star forming galaxies, the
  SFR-Stellar mass relationship, and the cosmic SFR density
  measurements, the basic observables used to compute the redshift 0
  point.

In this figure, we also attempt to relax the assumption of a constant
stellar mass function, and we use for each redshift shown the
mass functions derived for this redshift by \cite{ilbert09} (open
circles) and \cite{arnouts07} (open squares) in place of the AMF. The
general trend is similar to the one obtained with the AMF, but
introducing deviations around it. These deviations correspond to the
scatter in the observed stellar   mass functions. The approach consisting in
using the AMF allows us to focus on general trends, and filters out the
peculiarities of individual studies (e.g.  the effects of cosmic
variance).

The points at redshift larger than 1 can be considered as 
extrapolations in which we have little confidence since
the models are less and less reliable when moving at
larger redshifts. The points show that our basic assumptions
overpredict the cosmic SFR density  at redshift 1.75
by a few tens of dex. Since the models underpredict
the individual SFR in galaxies towards these redshifts but the
cosmic SFR density  is over-predicted, this means that our other assumption
(the consistency of the SMF) cannot be applied at these redshifts. Indeed,
the points computed with the SMF of \cite{ilbert09} and \cite{arnouts07}
are in better agreement with the \cite{wilkins08} data.

\begin{figure}
  \centering
  \includegraphics[width=9cm]{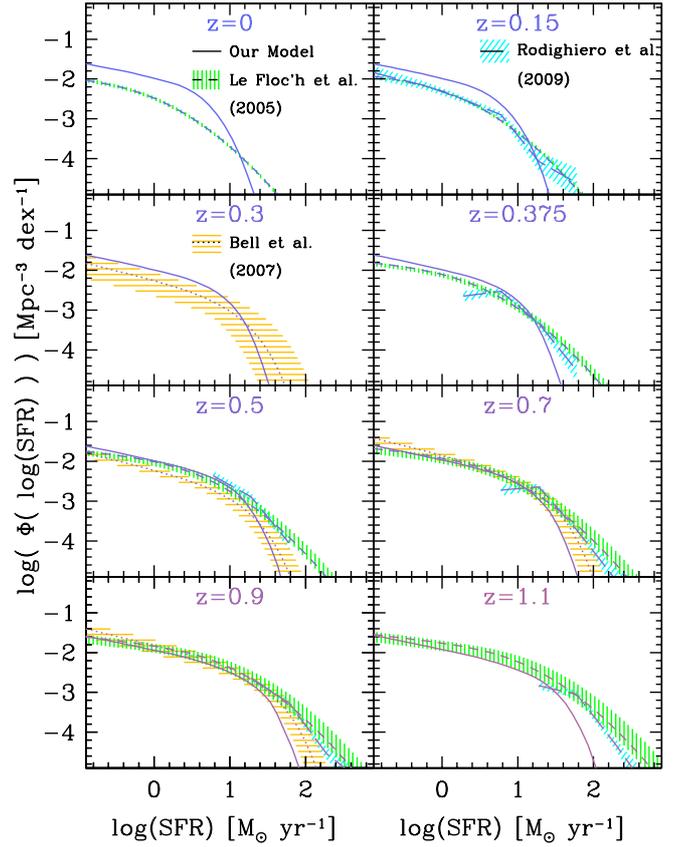}
  \caption{Models SFR Functions at various redshifts compared to
    observations of Bell et al.(2007) and to the infrared luminosity
    functions observed by \cite{lefloch05}, and \cite{rodighiero09} converted
    into SFR functions assuming $log(SFR) = log(L/L_{\odot})-9.97$
    (Buat et al. 08). }
  \label{FigSfrFunc}
\end{figure}

A good way to understand the cosmic SFR density evolution is
to look in Fig. \ref{FigSfrFunc} at the star forming galaxies
SFR functions at each redshift for which they were observationally
derived in Bell et al. (2007), \cite{lefloch05}, and
\cite{rodighiero09}.  The infra-red luminosity functions from the two
last papers were converted into SFR functions assuming $log(SFR) =
log(L/L_{\odot})-9.97$ (Buat et al. 08).  Note that other references
are available in the infrared but we show \cite{rodighiero09} and
\cite{lefloch05} for convenience as they are given for the same
redshift bins, of interest for us.  The latter study found a slightly
smaller number of very luminous galaxies than the former, as did
\cite{magnelli09}.
For redshifts above 0.3, the modeled SFR functions are consistent with the data, 
especially for intermediate star formation rates (a few to a few tens \msun{} yr$^{-1}$) 
dominating the cosmic SFR density  (see Fig. \ref{FigMultiEvo}). 
The shape of the function is not evolving much, but is 
shifted towards higher SFR when moving to larger redshifts, 
both in the observations and the models (as normal disk galaxies had
larger amount of gas and SFR in the past). 
The models distribution has a sharper drop at high SFR than the
observations (especially at high redshift), what may indicate that we
are missing a small number of galaxies with enhanced activity with
respect to the models.  This is also the case when comparing our
models to \cite{ilbert09} or \cite{pozzetti09}. We do not reproduce
the ``active'' fraction of galaxies observed by these studies.
However, because of the shape of the SFR function, such galaxies
cannot dominate the cosmic SFR density , although they may
contribute slightly to its evolution.
To estimate this contribution, we use the \cite{lefloch05}
  data to compute the fraction of the cosmic SFR density coming from
  galaxies with SFR higher than 10 $M_{\odot} yr^{-1}$ and that are
  not predicted by the models (i.e. computing the difference between
  the model and observed contributions to the cosmic SFR density above
  this SFR with respect to the total one).  This fraction is equal to
  13, 19, 23, 31 and 38 \% at redshift 0.375, 0.5, 0.7, 0.9, and 1.1.
  This discrepancy was found in other simple models (e.g. Daddi
  et al.  2007, Dav\'e et al.  2008). Part of the difference could be
  due to an enhancement of star formation during interactions that are
  not included in our approach.
Actually, the recent studies of \cite{robaina09} and \cite{jogee09}
suggest that major interactions trigger extra star formation. They 
contribute to less than 10 \% to the cosmic SFR density, but
are responsible for some of the galaxies with the largest SFR as revealed by
their infra-red luminosity. 
This results is also recovered by the empirical models of
\cite{hopkins09}.
Even so, our approach is still
meaningful since these galaxies do not represent the bulk of the contribution 
to the cosmic SFR density.

We have seen above that the model is on the upper side of the cosmic
SFR density at low redshift, thus predicting an evolution
with redshift on the weaker side of the one allowed by the data.  It
is interesting to note that the modeled SFR functions at low
redshift ($z<$0.3 panels in Fig. \ref{FigSfrFunc}) also overestimate
the number of galaxies with intermediate SFR, around a few $M_{\odot}
yr^{-1}$ (dominating the cosmic SFR density at this redshift)
by up to 0.6 dex (almost a factor 4). This difference explains the
excess in the cosmic SFR density obtained at low redshift in
the models.  The Stellar Mass-SFR relationship is relatively easy to
construct and different data sets provide very similar results for
this amount of star formation (top left panel of Fig.
\ref{FigDataSFRvsM}: the average values are well within 0.2 dex of
each other). The only other way to overestimate the cosmic SFR
density and the SFR function in our approach is to
over-estimate the Stellar Mass Function.  The AMF that we adopted is
however very close to the one measured by \cite{bell03} at redshift 0.
The most simple solution to this conundrum would be to consider that
the local stellar Mass Function reported in \cite{bell03} for
star forming galaxies overestimates the number of galaxies with
intermediate stellar mass (a few 10$^{10} M_{\odot} yr^{-1}$) by a
factor 2 to 4; and thus that the SMF in fact does evolve slightly at
the lowest redshifts (note that reducing the SMF for all redshifts
would lead us to underestimate the cosmic SFR density at
redshift $\sim$ 0.5, thus we do need an evolution).  We note that the
total observed stellar mass function of \cite{bell03} is in
good agreement with many other works, and is very probably fine.
However, there is an intrinsic difficulty in splitting galaxies in
star forming and quiescent ones (see the dispersion at larger redshift
between the various studies using various criteria for doing such a
thing).  Local studies are also very different in nature than higher
redshift ones (local volume versus deep field).  In summary, a small
evolution of the SMF of star-forming galaxies at the lowest redshift
could improve the agreement with SFR functions and the drop in the
cosmic SFR density.  Such an evolution is not visible in
current data, what may be due to the combined effect of dispersion and
of the intrinsic differences in local studies versus high redshift ones in
terms of surveys and separation of the various types of galaxies.

\subsection{On the fate of the Milky Way siblings}

\begin{figure}
  \centering
  \includegraphics[width=9cm]{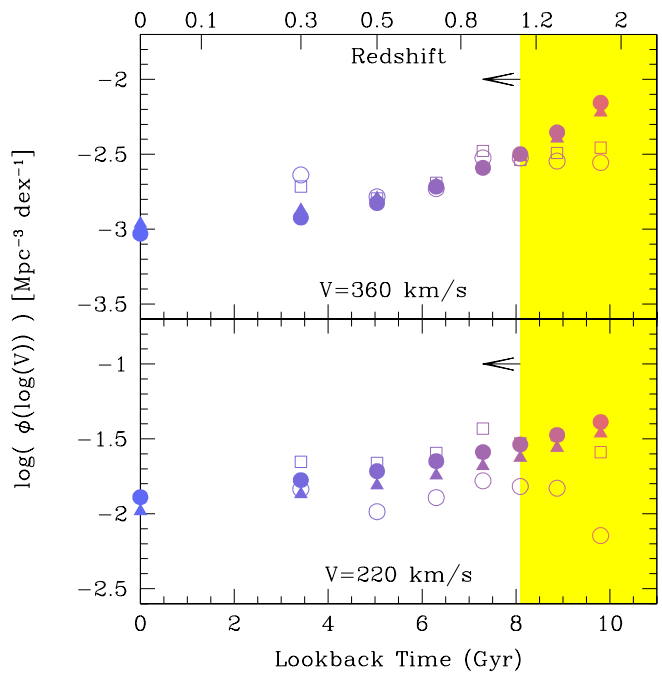}
  \caption{Evolution of the number of Milky-Way type spirals (V=220
    km/s, bottom) and more massive galaxies (V=360 km/s, top) with
    redshift. Symbols are the same as in Fig. \ref{FigCosmicSFR}.}
  \label{FigNumberEvoVgiven}
\end{figure}

Fig. \ref{FigNumberEvoVgiven} shows the number evolution of 
star forming galaxies of a given velocity. This is different from number 
evolution of galaxies of a given  stellar mass (what is more often shown) 
because the mass of a star forming galaxies evolves with time 
(e.g. Noeske 2009). In our approach, the velocity is attached 
to a galaxy. We show it for massive galaxies (V=360 km/s, top panel) and for 
intermediate mass galaxies, similar to the Milky Way (V=220 km/s, bottom panel). 
This figure shows clearly that the number of galaxies with a given 
velocity has declined since redshift 1.
The $V=220$ km/s galaxies have declined by a factor about 2 since
redshift 1. Thus statistically, half of the Milky Way siblings at
redshift 1 have disappeared since then ! The fraction of more massive
galaxies having survived is even lower.
Note that given the discussion above, these numbers are probably 
lower limits: if as suggested by the SFR function, the SMF 
does evolve, and especially is lower by a factor 2-4 at redshift 
zero, the evolution given above should be multiplied by a similar quantity.

Beyond redshift $\sim$ 1, once again our results are extrapolations
that should be treated with caution. Especially, the trend obtained
with a constant stellar Mass Function for star forming galaxies, and the one adopting
observed functions start to differ (this is similar to what is found
with the cosmic SFR density in the previous section). The
number of galaxies with $V=220 km/s$ is found to decline with
increasing redshifts when we use the  stellar Mass Functions from
\cite{ilbert09} (and flatten with Arnouts et al. 2007), indicating
that we may reach another era when the  stellar Mass Function of star forming
galaxies is not constant with time, as this population is building up.

What are the processes and loci of the transformation of half of the 
Milky Way sisters between redshift 1 and 0 ? Many suggestions
can be found in the literature.
For instance, \cite{pozzetti09} propose that the transformation 
could be driven by several processes (aging of the stellar 
population, exhaustion of the gas reservoir, gas stripping, 
AGN feedback, truncation of gas accretion via infall or satellites, 
starvation or strangulation) but they
do not clearly identify it.
\cite{bolzonella09} find some clues that the transformation 
is accelerated in over-dense regions. However the environment 
is only roughly defined (5th neighbor). They propose viable 
mechanism as merging processes triggering AGNs (but see 
discussion on mergers below) and propose finally strangulation 
(halo-gas stripping due to gravitational interaction with 
e.g. groups haloes). On the other hand, \cite{cowie08} 
found an effect of environment (traced
by the projected nearest neighbor) only
in massive galaxies (above 10$^{11}$ \msun), while we definitively 
need an effect below.
\cite{martig09} proposed the process of ``Morphological Quenching'' of the star
formation, occurring when disks become dominated by a stellar spheroid stabilizing
the disk against star formation. In this scenario, the spheroid can be built via major, 
minor merger or disk instabilities. Although an interesting 
process, this does not allow by itself
to easily predict which fraction of disks are going to turn 
red and dead.

Whatever process is responsible for quenching star formation, it cannot 
have happened only in clusters of galaxies that encompass a much smaller 
fraction of galaxies than the one we need to quench
(point also noted in Bolzonella et al. 2009).
Based on merger models, \cite{koda09} suggest that 70 \% of 
galaxies with halo mass in the range 5 $\times$ 10$^{10}$ - 10$^{12}$ \msun{}
have not suffered a merger with mass ratio larger than 
0.05 since redshift 1. This point is also consistent with
the results of \cite{oesch09} indicating that mergers 
cannot drive the morphological transformation at redshift 
less than unity. Observationally, \cite{robaina09} 
and \cite{jogee09} found relatively low rates of major mergers.
Thus major mergers cannot explain the number decrease of Milky Way 
galaxies (50\%) or more massive galaxies.

In conclusion, this major event in the history of a downsizing universe, 
consisting in the halt of star formation in previously star forming galaxies 
(then becoming quiescent galaxies) must be a common phenomenom occurring on an 
intermediate scale, not specifically related to clusters or major mergers. 
May be we should look at the scales of groups to find the culprit 
(see also Tran et al. 2009 and references within).
Several elements reinforce us in adopting groups as prime suspects, firstly the fact that
the number of galaxies in groups (55\% in Eke et al. 2004) is large enough to be 
consistent with the large fraction of galaxies that must quench their star formation.
More-over, Star Formation is suppressed on the scales of groups according to \cite{lewis02} who found that the SFR depends on local density, regardless of the richness of groups or clusters.
\cite{kilborn09} recently studied HI in groups and conclude that transformation is taking place in groups in which galaxies loose gas through tidal stripping during galaxy-galaxy or galaxy-group interaction.
Tanaka et al. (2007, 2009) also suggest that galaxy-galaxy interactions in groups play
a major role in the built-up of the red sequence on the basis of a detailed spectroscopic
analysis of galaxies endebbed in filaments around two clusters at redshift 0.55 and 1.24. 
Finally, Gavazzi et al. (work in preparation) found that the 
red sequence is already formed in groups (as it is in clusters, 
but not among isolated galaxies).

\section{Conclusions}
\label{SecConclu}
This paper presents an original and educational approach to work with
Stellar Mass Function of star forming galaxies. The main point
consists in combining observed Stellar Mass Functions with simple
backward evolutionary models (assuming a smooth evolution driven by
star formation in a galactic disk fuelled by cold accretion) to obtain
modeled velocity functions at various redshifts. The big
advantage of the velocity is that it is a model parameter attached to
a galaxy, contrary to stellar mass, which evolves with star formation.
In the framework of such models, for a given velocity, the Star
Formation History and the Stellar Mass history are known. Galaxies
with large velocities are the precursor of today's massive star
forming galaxies (in the absence of quenching), with stars having
formed at an early epoch. Galaxies with low velocities have been formed on
average at later times. These histories are not adhoc models for this
paper, but have been shown to be consistent with a large number of
properties of nearby star-forming galaxies.  Combining the Velocity
Functions and the models, it is possible to compute the model
  predictions for a large number of quantities and distributions at
any redshift (e.g. SFR functions and cosmic SFR density).

At the current time, the empirical  stellar mass functions are still rather 
uncertain (see the scatter between various studies). 
Assuming that, to the first order, the stellar mass function is constant for star 
forming galaxies at redshift lower than 1 (in accordance with
the data), we obtained several interesting results:

   \begin{enumerate}

   \item A globally consistent picture is obtained in which a large
     fraction of star forming galaxies quench their activity of star
     formation between redshift 0 and 1, building up the red sequence.
     The shape of the quiescent galaxies stellar mass
     function is recovered above $\sim$ 3 10$^{10}$
       \msun.  Mergers can only alter it slightly by
     redistributing the mass among galaxies. At most, their
     main effect could be seen only on the massive end of the
     stellar Mass Function.

\item The evolution of the cosmic SFR density falls within
  the compilation of \cite{wilkins08}. However, our decrease towards
  low redshifts is smaller than in the observations.
The discrepency is especially due to our redshift 0 point. This may
be related to a small inconsistency between the stellar mass function,
SFR function, and stellar mass-SFR relationship of the local
  universe star forming galaxies.

\item The fraction of galaxies having quenched their star formation is
  high, in rough agreement with measurements.  For example, we
  conclude that half of the Milky Way siblings (galaxies with
    circular velocities of 220 km/s) have quenched their star
  formation since redshift 1, and are then no more star-forming disk
  such as the Galaxy.  This fraction is probably a lower limit since
  the model cosmic SFR density presents itself a relatively
    weak evolution with respect to the observed one.

\item The process responsible for this change is not clearly identified today. 
What is clear is that it is probably not related to cluster physics 
or to major mergers (which affect galaxies in too small numbers). 
Groups are  our prime suspects because they are numerous enough to have 
a significant effect on the cosmic evolutions and several studies indicate 
that transformations are indeed taking place in them.

   \end{enumerate}

In the future, we hope that wider, deeper fields and improvement in the 
analysis of stellar mass functions will allow us to use the method 
exposed in this paper but on safer grounds. This will allow us to improve
our models of galaxy evolution and to get a more
complete picture of the  cosmic evolution of star-forming  galaxies, 
and their transformation into quiescent ones.

\begin{acknowledgements}
  We acknowledge a long history of interactions with Nikos Prantzos on
  the subject of modeling the evolution of galaxies, and we thank
  St\'ephane Arnouts, Alessandro Boselli, Luca Cortese and Kevin Bundy for discussions
  concerning stellar mass functions, quenching and
  environmental effects.  We also thank Xiao Liang Liu for
  interactions in the early phases of this work, 
  and S\'ebastien Heinis for carefully reading our manuscript.
\end{acknowledgements}

\end{document}